\documentclass[twocolumn,showpacs,preprintnumbers,amsmath,amssymb,aps]{revtex4}
\usepackage{graphicx}
\usepackage{dcolumn}
\usepackage{bm}
\begin{document}
\bibliographystyle{apsrev}

\title{Ground state of a confined Yukawa plasma including correlation effects}
\author{C.~Henning$^1$}
\author{P.~Ludwig$^{1,2}$}
\author{A.~Filinov$^1$}
\author{A.~Piel$^3$}
\author{M.~Bonitz$^1$}
\email{bonitz@physik.uni-kiel.de}
\affiliation{$^1$Institut f\"ur Theoretische Physik und Astrophysik, Christian-Albrechts-Universit\"{a}t zu Kiel, D-24118 Kiel, Germany}
\affiliation{$^2$Institut f\"ur Physik, Universit\"{a}t Rostock, Universit\"{a}tsplatz 3, D-18051 Rostock, Germany}
\affiliation{$^3$Institut f\"ur Experimentelle und Angewandte Physik, Christian-Albrechts-Universit\"{a}t zu Kiel, D-24118 Kiel, Germany}

\date{\today}

\begin{abstract}
The ground state of an externally confined one-component Yukawa plasma is derived analytically using the local density approximation (LDA). In particular, the radial density profile is computed. The results are compared with the recently obtained mean-field (MF) density profile \cite{henning.pre06}. While the MF results are more accurate for weak screening, LDA with correlations included yields the proper description for large screening. By comparison with first-principle simulations for three-dimensional spherical Yukawa crystals we demonstrate that both approximations complement each other. Together they accurately describe the density profile in the full range of screening parameters.
\end{abstract}
\maketitle

\section{Introduction}
Interacting particles in confinement potentials are omnipresent in nature and laboratory systems such as trapped ions, e.g. \cite{itano,drewsen}, dusty plasmas, e.g. \cite{goree,zuzic,hayashi} or ultracold Bose and Fermi gases \cite{ohashi,pethicksmith}. An interesting aspect of particle traps is that it is easy to realize situations of strong correlations. The observed particle arrangements reach from gas-like, liquid-like to solid behavior where the symmetry is influenced by the trap geometry. Of particular recent interest have been spherical traps in which plasma crystals consisting of spherical shells (Yukawa balls) are formed, e.g. \cite{arp04,totsuji05,ludwig-etal.05pre,bonitz-etal.prl06}. The particle distribution among the shells is by now well understood \cite{bonitz-etal.prl06,golubnychiy.jp06,baumgartner.cpp07}.

In a recent study \cite{henning.pre06} we have analyzed also the average particle density in the trap and found that it is very sensitive to the binary interaction: it changes from a flat profile in case of long-range Coulomb interaction to a profile rapidly decaying away from the trap center in the case of a screened Yukawa potential. Using a non-local mean-field (MF) approximation the density profile could be computed analytically and was found to agree very well with first principle computer simulations for Yukawa crystals. However, when the screening is increased deviations in the trap center kept growing which were attributed to correlation effects missing in the mean-field model.

The goal of this paper is to remove these discrepancies. For this we extend the analysis of Ref.~\cite{henning.pre06} by including correlation effects following an idea of Totsuji et al. \cite{totsuji01} applied to 2D systems. We apply the local density approximation (LDA) using known results \cite{totsuji.jp06} for the correlation energy of a homogeneous one-component Yukawa plasma. The results clearly confirm that correlation effects are responsible for the strong density increase in the trap center. We find that LDA with correlations included agrees very well with simulations of Yukawa crystals in the limit of strong screening. On the other hand, for weak screening, the previous MF result turns out to be more accurate. Interestingly, for intermediate values of the screening parameter both methods are accurate, so a combination of both allows to quantitatively describe the density profile in the whole range of screening parameters.

This paper is organized as follows. In Sec. II we introduce the LDA and use it first to compute the density profile in a mean-field approximation, which, of course, gives worse results than MF, but helps to understand LDA. Then, in Sec. III we improve the LDA model by including correlation effects. In Sec. IV the results for the density profile are compared to molecular dynamics simulations. A discussion is given in Sec. V.

\section{Ground state of a confined plasma within LDA}
We consider $N$ identical particles with mass $m$ and charge $Q$ confined by an external potential $\Phi$ and interacting with the isotropic Yukawa-type pair potential, $V(r)=(Q^2/r)\exp(-\kappa r)$. To derive the properties of interest we start with the expression of the ground-state energy, which is given by
\begin{equation}\label{eq:energy}
	E[n]=\int d^3r\, u({\bf r}),
\end{equation}
with the energy density $u({\bf r})=u^\text{mf}({\bf r}) + u^\text{cor}({\bf r})$, where the mean-field energy density is
\begin{equation}\label{eq:energy_density}
	u^\text{mf}({\bf r})=n({\bf r})\biggl\{\Phi({\bf r})+\frac{N-1}{2N}\int d^3r_2\,n({\bf r}_2)V(|{\bf r}-{\bf r}_2|)\biggr\}.
\end{equation}
The correlation contribution $u^\text{cor}$ will be discussed below (Sec. III) by means of the local density approximation. Before, we introduce this approximation and obtain first the LDA results in mean-field approximation (LDA-MF). These results will not be as accurate as the MF results, due to the applied approximation, but LDA-MF helps to familiarize oneself with LDA and its characteristics.

The local density approximation is based upon the idea of replacing the non-local terms within the energy density at point ${\bf r}$ by local expressions using the known energy density of the homogeneous system with its density $n_0$ equal to the local density $n({\bf r})$ of the true inhomogeneous system in question. Therefore, to derive LDA-MF we need to substitute for the non-local second term in \eqref{eq:energy_density}, i.e. for the density of interaction energy, the corresponding expression of the infinite homogeneous system, which is given by [details are given in the Appendix]
\begin{align}\label{eq:energy_density_homogsystem}
	u_0(\kappa)&=n_0\frac{N-1}{2N}Q^2\int d^3r_2\,n_0 \frac{e^{-\kappa|{\bf r}-{\bf r}_2|}}{|{\bf r}-{\bf r}_2|}\\
	&=\frac{N-1}{N}Q^2\,n_0^2 \frac{2\pi}{\kappa^2},\notag
\end{align}
and, as a second step, replace the homogeneous density $n_0$ by the local density $n({\bf r})$. Thus we obtain the LDA-MF ground-state energy
\begin{equation}\label{eq:energy_LDA}
  	E^\text{mf}_\text{LDA}[n]=\int d^3r\, u({\bf r})
\end{equation}
with the energy density
\begin{equation}\label{eq:energydensity_LDA}
	u({\bf r})=n({\bf r})\biggl\{\Phi({\bf r})+\frac{N-1}{N}Q^2\,n({\bf r})\frac{2\pi}{\kappa^2}\biggr\}.
\end{equation}
The variation of the energy
\begin{equation}\label{eq:extendedvar_problem}
	\tilde{E}^\text{mf}_\text{LDA}[n,\mu]=E^\text{mf}_\text{LDA}[n]+\mu\biggl\{N-\int d^3r\, n({\bf r})\biggr\}
\end{equation}
with respect to the density $n({\bf r})$ (for details see Ref.~\cite{henning.pre06}) yields an explicit expression for the density profile in an arbitrary confinement potential
\begin{equation}\label{eq:general_LDAdensity}
	n({\bf r})=\frac{N\kappa^2}{4\pi(N-1)Q^2}\Bigl(\mu-\Phi({\bf r})\Bigr),
\end{equation}
which holds at any point where the density is nonzero. Due to \eqref{eq:extendedvar_problem} this density is normalized by
\begin{equation}\label{eq:normalization}
	\int d^3r \,n({\bf r})=N.
\end{equation}

The case of an isotropic confinement $\Phi({\bf r})=\Phi(r)$, which is of particular interest, leads to an isotropic density distribution $n({\bf r})=n(r)={\tilde n}(r)\Theta(R-r)$ the outer radius $R$ of which is being fixed by the normalization condition \eqref{eq:normalization} which now becomes $\int_0^R dr\, r^2 {\tilde n}(r)=N/4\pi$. In this isotropic case the yet unknown Lagrange multiplier $\mu$ can be obtained by taking the variation also with respect to $R$ \cite{totsuji01}, which yields
\begin{equation}\label{eq:general_LDAmu}
  \mu=\Phi(R).
\end{equation}

Compared to the MF result which was given in \cite{henning.pre06},
\begin{gather}\label{eq:general_meanfielddensity}
	n^{\text{mf}}({\bf r})=\frac{N}{4\pi(N-1)Q^2}\Bigl(\Delta\Phi({\bf r})+\kappa^2\mu^{\text{mf}}-\kappa^2\Phi({\bf r})\Bigr),\\
	\mu^{\text{mf}}=\Phi(R^{\text{mf}})+\frac{R^{\text{mf}}\,\Phi'(R^{\text{mf}})}{1+\kappa R^{\text{mf}}}\label{eq:general_meanfieldmu},
\end{gather}
the LDA-MF density \eqref{eq:general_LDAdensity} differs in two points. On the one hand the Laplacian of the potential $\Delta\Phi({\bf r})$ is missing and, on the other hand, the expression of the chemical potential $\mu$ is simpler than $\mu^{\text{mf}}$. That is based upon the fact, that the missing expressions consist of derivatives and thus contain informations about contiguous values of the potential, what is suppressed within LDA-MF and generally within LDA.

\subsection{Parabolic confinement potential}
\begin{figure}
\includegraphics[width=8.5714cm,height=6cm,clip=true]{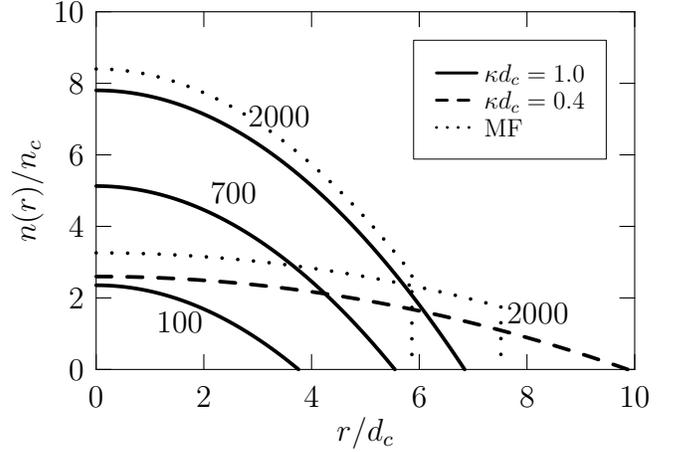}
\caption{Radial density profile for a parabolic confinement potential $\Phi(r)=(\alpha/2) r^2$ and a constant screening parameter $\kappa d_c=1$ and three different particle numbers $N=100, 700, 2000$. Also the result for $\kappa d_c=0.4$, $N=2000$ is shown by the dashed line. For comparison, the non-local MF results for $\kappa d_c=0.4,1.0$, $N=2000$ are given by the dotted lines.}
\label{fig:LDAresults}
\end{figure}
For the case of a parabolic external potential $\Phi(r)=(\alpha/2) r^2$ the density following from Eqs.~\eqref{eq:general_LDAdensity} and \eqref{eq:general_LDAmu} is
\begin{equation}\label{eq:parabolic_LDAdensity}
	n(r)=\frac{\alpha N}{4\pi (N-1) Q^2}\left(\frac{\kappa^2 R^2}{2}-\frac{\kappa^2r^2}{2}\right)\Theta(R-r).
\end{equation}
The dimensionless combination $\kappa R$, which contains the limiting outer radius, can be obtained from the normalization \eqref{eq:normalization} and is given by
\begin{equation}\label{eq:LDAradius}
	\begin{split}
		\kappa R&=\sqrt[5]{\frac{15 (N-1) Q^2 \kappa^3}{\alpha}}\\
		 &=\sqrt[5]{\frac{15}{2}(\kappa d_c)^3(N-1)}.
	\end{split}
\end{equation}
Here, we introduced the length scale $d_c=(2Q^2/\alpha)^{1/3}$, which is the stable distance of two charged particles in the absence of screening \cite{bonitz-etal.prl06} and which will be used below as the proper unit for lengths and screening parameters. As unit for densities we use the average density of a large Coulomb system, which is given by $n_c=(3\alpha)/(4\pi Q^2)$.

The results of \eqref{eq:parabolic_LDAdensity} are shown in Fig.~\ref{fig:LDAresults} for three particle numbers from $N=100$ to $N=2000$. One clearly sees the parabolic decrease of the density away from the trap center till it terminates in zero. The curvature of the density does not change by increasing the particle number - just the density increases continuously at every space point and, at the same time, extends to higher values of the limiting radius $R$. However, the curvature of the density profile changes dramatically, when the plasma screening is increased at constant $N$.

Thus, in the case of an isotropic parabolic potential, the LDA density profile bears qualitative resemblance to the density profile in the non-local mean-field approximation, although in the case of other confinement potentials the deviations of the LDA-MF from the MF approximation are stronger, cf. Eqs.~\eqref{eq:general_LDAdensity} and \eqref{eq:general_meanfielddensity}. However, quantitatively in two points MF differs from LDA-MF for parabolic confinement as well, as can also be seen in Fig.~\ref{fig:LDAresults}.
\begin{figure}
\includegraphics[width=8.5714cm,height=6cm,clip=true]{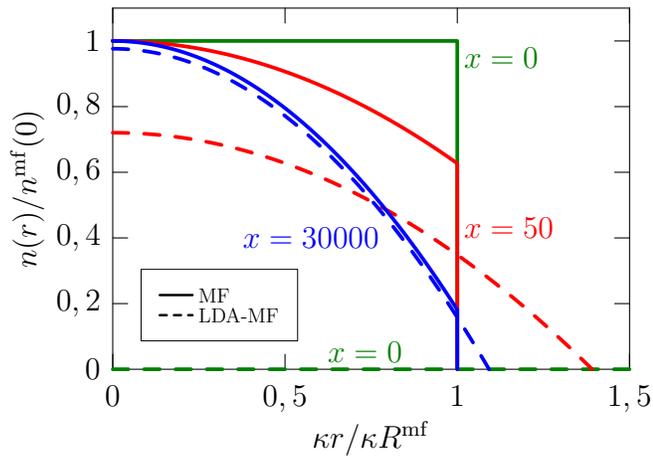}
\caption{(Color online) Radial MF density profile (solid lines) compared to LDA-MF (dashed lines) for three different density parameters $x=(\kappa d_c)^3(N-1)$. The abscissa is normalized with the MF radius $R^\text{mf}$, while the ordinate is normalized with the corresponding MF density $n^\text{mf}(0)$ at the trap center.}
\label{fig:LDAcomparison}
\end{figure}
First, the density in this local density approximation does not show a discontinuity at $r=R$, in contrast to the MF result, Eqs.~(\ref{eq:general_meanfielddensity}, \ref{eq:general_meanfieldmu}). This is due to the neglect of edge effects in this derivation of the LDA result. Secondly, LDA-MF yields too small values of the density. This error is reduced (cf. Fig.~\ref{fig:LDAcomparison}) with increasing values of the density parameter $x=(\kappa d_c)^3(N-1)$, cf. Ref.~\cite{henning.pre06}, which regardless of the factor $N/(N-1)$ solely determines the density profile. The reason for this improved behavior with increasing $x$ is due to the fact that an increase of $\kappa$ contracts the effective area of integration within \eqref{eq:energy_density} as well as within \eqref{eq:energy_density_homogsystem}. The contraction finally is in favor of the accuracy of LDA-MF, because the decreased integration volume contains a more homogeneous density. Also an increase of the particle number $N$, what flattens the density profile, will similarly improve LDA-MF.

Because the validity of the mean-field model depends on the value of the screening parameter $\kappa d_c$, there are the following two cases. In the first case, for small values of the screening parameter, the MF approximation provides a good description of the density profile, but LDA-MF underrates this profile and so does not give a good description on its own. (That applies also if finite-size effects are included, cf. Fig.~\ref{fig:LDA-finitesize}.) In the second case, for large values of the screening parameter, the LDA-MF approaches the MF approximation, however, there, the latter does not describe the density profile correctly due to the neglect of the now relevant correlation contributions \cite{totsuji01}. Thus, the local density approximation of the mean-field energy alone does not give a good description of the density profile.

However, it gives a straightforward way to include the missing correlation contributions in the energy density by usage of the result of the homogeneous system, see Sec. \ref{section_correlation}.

\subsection{Improvement of LDA by inclusion of finite-size effects}\label{section_fs_correction}
As can be seen from Fig.~\ref{fig:LDAcomparison} and from Eq.~\eqref{eq:general_LDAdensity} the density profile obtained by LDA-MF breaks down in the Coulomb case - the density cannot be normalized anymore, which is the same as in the two-dimensional case \cite{totsuji01}. But the application of a local density approximation cannot be the reason for this, because the method of LDA is based upon the usage of results from the homogeneous system, and the Coulomb system is homogeneous with $n_0=[N/(N-1)] n_c$.
\begin{figure}
\includegraphics[width=8.5714cm,height=6cm,clip=true]{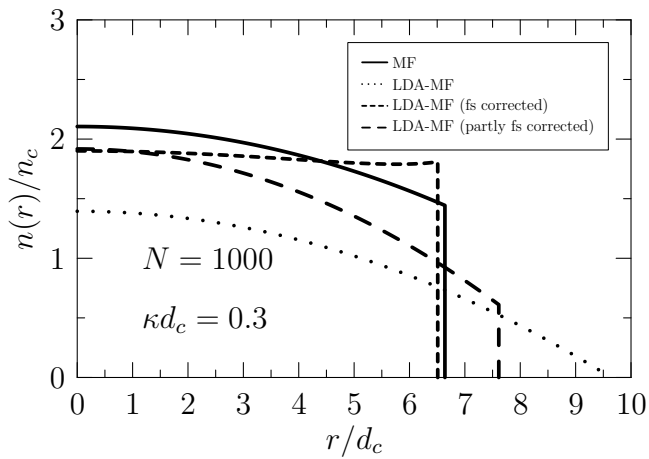}
\caption{Radial density profiles of a spherical plasma of $N=1000$ and $\kappa d_c=0.3$ calculated by LDA-MF with (fs corrected) and without finite-size effects included. For comparison the exact MF result is also given by the dashed line. The difference between the finite-size correction and the partial finite-size correction is described in the text.}
\label{fig:LDA-finitesize}
\end{figure}

In fact, the cause of the breakdown is the use of results from the infinite homogeneous system neglecting finite-size effects. This failure can be avoided by replacing \eqref{eq:energy_density_homogsystem} by the corresponding expression of the finite homogeneous system. In the appendix such an expression is derived for isotropic confinement. As a result the finite-size effects lead to a corrected density profile
\begin{equation}\label{eq:parabolic_LDAdensity_fscorrected}
\begin{split}
	n(r)=&\frac{N\kappa^2}{4\pi(N-1)Q^2}\\
	&\times\frac{\mu-\Phi(r)}{1-e^{-\kappa R}(1+\kappa R)\sinh(\kappa r)/(\kappa r)}\Theta(R-r),
\end{split}
\end{equation}
instead of Eq.~\eqref{eq:parabolic_LDAdensity}, what indeed yields the constant MF solution also for LDA-MF. As another example, in Fig.~\ref{fig:LDA-finitesize} the density profiles with [LDA-MF (fs corrected)] and without these finite-size contributions are shown for $N=1000$, $\kappa d_c=0.3$. One clearly sees, that in the case of finite-size correction the density profile shows a discontinuity at the boundary and, due to that, it yields increased values of the density. However, the density profile including edge effects is not monotonically decreasing away from the trap center but has a density increasing part in the outer range, what is not correct. This is due to the spatial dependence of the denominator of Eq.~\eqref{eq:parabolic_LDAdensity_fscorrected}.

By contrast a more accurate monotonically decreasing density profile can also be obtained by taking the finite-size effects only partly into account [LDA-MF (partly fs corrected)], as derived in the appendix. The final result is given by
\begin{equation}
	n(r)=\frac{N\kappa^2}{4\pi(N-1)Q^2}\frac{\mu-\Phi(r)}{1-e^{-\kappa R}(1+\kappa R)}\Theta(R-r),
\end{equation}
which now misses the spatial dependence in the denominator. The corresponding result is also given in Fig.~\ref{fig:LDA-finitesize}.

Consequently, for Yukawa systems like those analyzed here an improvement of LDA is possible by including edge effects. However, for small values of the screening parameter even the improved local density approximation does not approach the degree of accuracy obtained by the non-local mean-field model, cf. Fig.~\ref{fig:LDA-finitesize}. On the other hand, for increased screening the finite-size effects do not alter the density profile significantly.

Therefore, below we continue to use Eq.~\eqref{eq:energy_density_homogsystem} of the infinite homogeneous system, what will not interfere the following results.

\section{Inclusion of correlation contributions}\label{section_correlation}
The energy expression $E^\text{mf}_\text{LDA}$ (\ref{eq:energy_LDA},\ref{eq:energydensity_LDA}) contains only the energy density of the confinement and of the mean-field interaction. To include the contribution of the particle correlations we can make use of the result for the density of correlation energy of the homogeneous system which is given by Eq.~(3) of Ref.~\cite{totsuji.jp06}
\begin{equation}
	\begin{split}
		u_\text{corr}(n_0,\kappa)=&-1.444 Q^2 n_0^{\frac{4}{3}}\\
		              &\times\exp\Bigl(-0.375\kappa n_0^{-\frac{1}{3}}+7.4\cdot10^{-5}(\kappa n_0^{-\frac{1}{3}})^4\Bigr),
	\end{split}
\end{equation}
where $n_0$ is the corresponding density of the homogeneous system. By replacing this density with the local density $n({\bf r})$ of the inhomogeneous system one obtains the correlation contribution of the energy density within LDA. Thus we derive the complete ground-state energy in local density approximation
\begin{equation}\label{eq:energyLDAcorr}
  	E_\text{LDA}[n]=\int d^3r\, u({\bf r})
\end{equation}
with energy density
\begin{equation}
	\begin{split}
		u({\bf r})=&n({\bf r})\Phi({\bf r})+\frac{N-1}{N}Q^2\,n({\bf r})^2\frac{2\pi}{\kappa^2}-1.444 Q^2 n({\bf r})^{\frac{4}{3}}\\
		           &\times\exp\Bigl(-0.375\kappa n({\bf r})^{-\frac{1}{3}}+7.4\cdot10^{-5}(\kappa n({\bf r})^{-\frac{1}{3}})^4\Bigr).
	\end{split}
\end{equation}
As before, variation of the energy \eqref{eq:energyLDAcorr} at constant particle number, cf. Eq.~\eqref{eq:extendedvar_problem}, yields the ground state density profile, but now with correlation effects included. In this case the strong non-linear character of the energy density does not allow for an explicit solution. Just an implicit solution is possible and is given by the following equation for $z^3({\bf r})=\kappa^{-3}n({\bf r})$, which can be regarded as the local plasma parameter of the system,
\begin{equation}\label{eq:LDAcorrdensity}
	\begin{split}
		0=&\frac{N-1}{N}z^3({\bf r})+\frac{\Phi({\bf r})-\mu}{4\pi Q^2 \kappa}-\bigl(c_1 z({\bf r}) + c_2 - c_3 z({\bf r})^{-3}\bigr)\\
		  &\times\exp\Bigl(-0.375 z({\bf r})^{-1}+7.4\cdot10^{-5}z({\bf r})^{-4}\Bigr).
	\end{split}
\end{equation}
The constants $c_i$ are given by
\begin{subequations}
\begin{align}
	c_1&=0.153\\
	c_2&=0.0144\\
	c_3&=1.134\cdot10^{-5}.
\end{align}
\end{subequations}

\begin{figure}
\includegraphics[width=8.5714cm,height=6cm,clip=true]{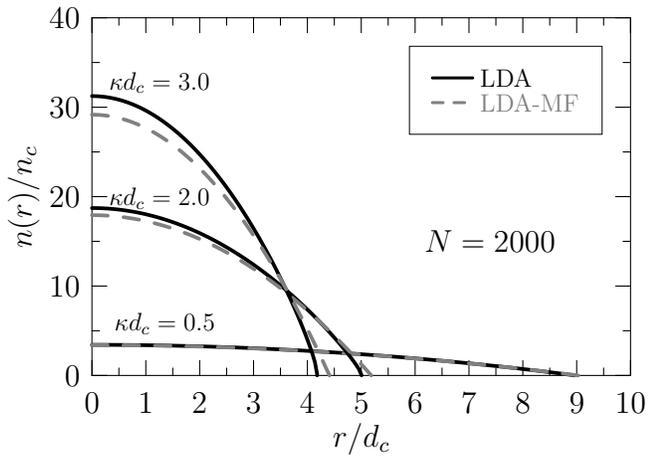}
\caption{Radial density profile of a confined spherical plasma of $N=2000$ particles calculated with LDA including correlation contributions (solid lines) compared to LDA-MF (dashed lines) for three different screening parameters.}
\label{fig:LDAcorrresults}
\end{figure}

The solution of Eq.~\eqref{eq:LDAcorrdensity} can be obtained numerically. For the case of a parabolic external potential $\Phi(r)=(\alpha/2) r^2$ results are given in Fig.~\ref{fig:LDAcorrresults}. There, the density profile of a plasma of $N=2000$ particles within LDA is shown for three different screening parameters: $\kappa d_c=0.5$, $\kappa d_c=2.0$ and $\kappa d_c=3.0$. For comparison the LDA-MF density profile is shown too.

\begin{figure}
\includegraphics[width=8.5714cm,height=6cm,clip=true]{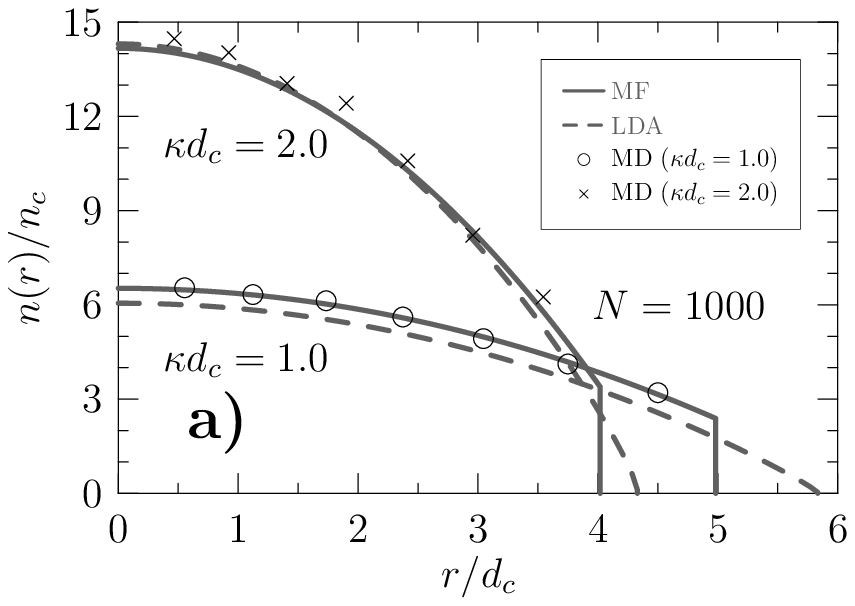}
\includegraphics[width=8.5714cm,height=6cm,clip=true]{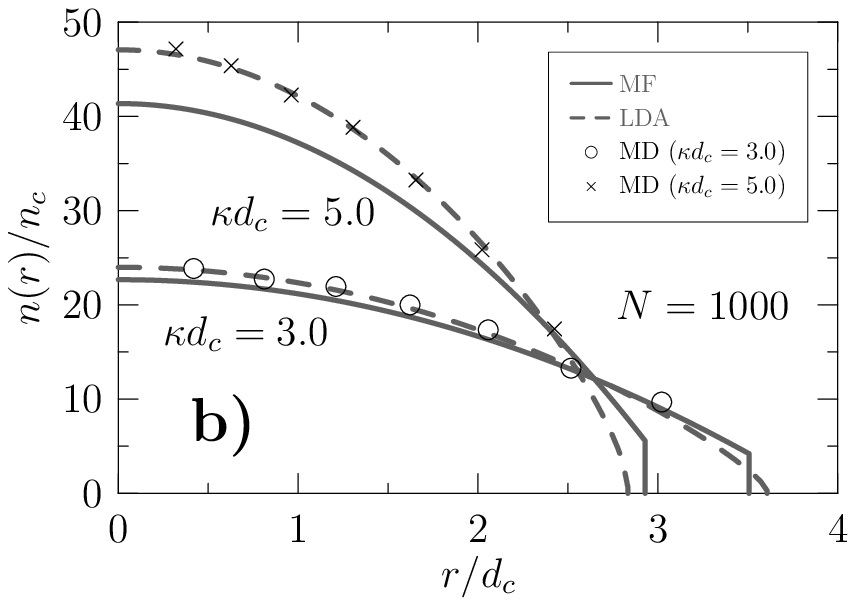}
\caption{Radial density profiles of a three-dimensional plasma of $N=1000$ particles calculated with
the exact mean-field model (solid lines) and with LDA including correlation contributions (dashed lines) for four different screening parameters: $\kappa d_c=1$, $\kappa d_c=2$, $\kappa d_c=3$ and $\kappa d_c=5$. Averaged shell densities of molecular dynamics results of a plasma crystal for the same parameters are shown by the symbols. For details see discussion in Sec.~\ref{sec:Comparison_with_simulation_results}. }
\label{fig:LDAcorrComparison}
\end{figure}

It can be seen that for a small screening parameter (see line $\kappa d_c=0.5$) both density profiles are nearly identical. But with increasing screening, i.e. for smaller values of the local plasma parameter $z^3$, the correlation contributions within LDA alter the curvature of the profile, which rises more steeply towards the center. So the particle correlations tend to increase the central density of the plasma, which also can be seen in Fig.~\ref{fig:LDAcorrComparison} in comparison with the mean-field approximation.
\begin{figure}
\includegraphics[width=8.5714cm,height=6cm,clip=true]{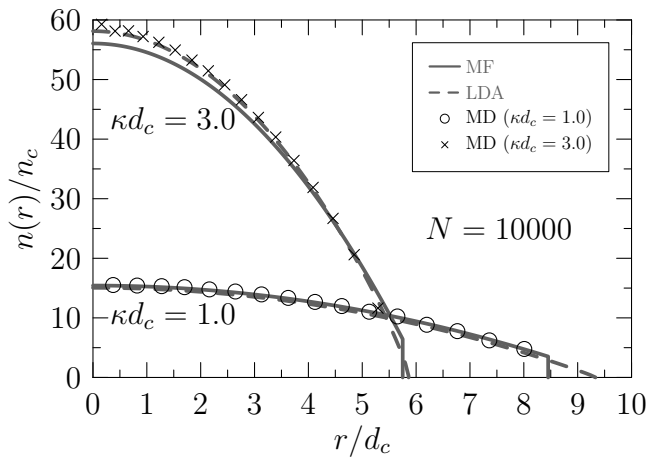}
\caption{Radial density profiles of a three-dimensional plasma with $N=10000$ and two different screening parameters ($\kappa d_c=1$, $\kappa d_c=3$). The solid (dashed) lines show MF (LDA) calculation results. Symbols denote molecular dynamics results of a plasma crystal for the same parameters where the average density at the positions of the shells is shown.}
\label{fig:LDAcorrComparison2}
\end{figure}
\section{Comparison with simulation results for finite Yukawa crystals}\label{sec:Comparison_with_simulation_results}
We performed molecular dynamics simulations of the ground state of a large number of Coulomb balls for the purpose of comparison with the average density of spherical Yukawa crystals, see refs.~\cite{ludwig-etal.05pre, bonitz-etal.prl06} for details. In order to obtain a smooth average radial density profile the averaging process was accomplished by substituting each particle by a small but finite sphere. Numerical results of this comparison with a Coulomb ball of $N=1000$ particles are included in Fig.~\ref{fig:LDAcorrComparison} for four different screening parameters. The symbols denote the average particle density in the vicinity of the corresponding shell, while the lines represent the MF (solid) and the LDA density (dashed).

For small values of the screening parameter $\kappa d_c<2$ the simulation results are very well reproduced by the analytical density profile of the non-local mean-field model (MF), whereas the local density approximation underrates the results (lower lines in Fig.~\ref{fig:LDAcorrComparison} (a)). On the other hand, for larger values of the screening parameter $\kappa d_c>2$ the simulation results are reproduced by LDA, whereas MF underestimates these results in the center. This underestimation is accompanied by a wrong prediction of the profile curvature (Fig.~\ref{fig:LDAcorrComparison} (b)). For intermediate values of the screening parameter $\kappa d_c\approx2$ both methods are very close to the averaged simulation results (upper lines in Fig.~\ref{fig:LDAcorrComparison} (a)). We have verified this behavior also for other Coulomb balls. Another representative example is shown in Fig.~\ref{fig:LDAcorrComparison2} for a Coulomb ball with $N=10000$. There, the same behavior as in Fig.~\ref{fig:LDAcorrComparison} is seen.

\section{Summary and discussion}
A theoretical analysis of the ground state density profile of a spatially confined one-component plasma within local density approximation was presented. We derived a closed equation, Eq.~\eqref{eq:LDAcorrdensity}, for the density profile including correlations effects for arbitrary confinement potentials with any symmetry. In contrast to the result without particle correlations the density profile shows an increased central density with increasing screening parameter. The validity of LDA is however limited to not too small values of the screening parameter, $\kappa d_c \ge 2$.

Comparisons with first-principle simulation results of strongly correlated Coulomb clusters with varying screening parameter showed that LDA allows to remove the problem of the MF approximation observed in Ref.~\cite{henning.pre06} which arises with increasing screening parameter. Therefore, the mean-field model together with the presented local density approximation complement one another in the description of strongly correlated spatially confined one-component plasmas.

\begin{acknowledgments}
The authors are indebted to D.~Block and A.~Melzer for fruitful discussions. This work is supported by the Deutsche Forschungsgemeinschaft via SFB-TR 24 projects A3, A5 and A7. 
\end{acknowledgments}

\appendix
\section{Local density approximation using a finite reference system}
The investigation of an inhomogeneous system within LDA uses known results from the corresponding homogeneous system. There, the infinite homogeneous system is often used as reference system what entails that finite-size effects are neglected. To take these into account the finite homogeneous system has to be used as reference. In the present derivation such a modification is made for an isotropic confinement and leads to a change of the expression for the density of interaction energy, Eq.~\eqref{eq:energy_density_homogsystem},
\begin{equation}
\begin{split}
	u_0(\kappa)&=n_0\frac{N-1}{2N}Q^2\int d^3r_2\,n_0 \frac{e^{-\kappa|{\bf r}-{\bf r}_2|}}{|{\bf r}-{\bf r}_2|}\\
	&=n_0^2\frac{N-1}{2N}Q^2\int d^3r_2\,4\pi r_2^2\frac{e^{-\kappa r_2}}{r_2}\\
	&=\frac{N-1}{N}Q^2\,n_0^2 \frac{2\pi}{\kappa^2}.
\end{split}
\end{equation}
This formula has no spatial dependence due to the infinite integration volume and it diverges in the limit of Coulomb interaction ($\kappa\rightarrow0$) leading to a breakdown of the approximation.

By contrast, the density of interaction energy of the corresponding finite homogeneous system (a sphere with center ${\bf r}_2=0$ and radius $R$) is given by
\begin{equation}
\begin{split}
	u_0&(\kappa,r)\\
	&=n_0\frac{N-1}{2N}Q^2\int_{\mathcal{K}(R,0)} d^3r_2\,n_0 \frac{e^{-\kappa|{\bf r}-{\bf r}_2|}}{|{\bf r}-{\bf r}_2|}\\
	&=n_0^2\frac{N-1}{2N}Q^2 \frac{2\pi}{\kappa r}\int_0^R dr_2\, r_2\Bigl\{-e^{-\kappa(r+r_2)}+e^{-\kappa|r-r_2|}\Bigr\}\\
	&=n_0^2\frac{N-1}{2N}Q^2 \frac{4\pi}{\kappa r}\biggl(e^{-\kappa r}\int_0^r dr_2\,r_2\sinh(\kappa r_2)\\
	&\hphantom{=n_0^2\frac{N-1}{N}Q^2 \frac{4\pi}{\kappa r}\biggl(}+\sinh(\kappa r)\int_r^R dr_2\, r_2 e^{-\kappa r_2}\biggr)\\
	&=\frac{N-1}{N}Q^2\,n_0^2 \frac{2\pi}{\kappa^2}\biggl(1-e^{-\kappa R}(1+\kappa R)\frac{\sinh(\kappa r)}{\kappa r}\biggr),
\end{split}
\end{equation}
including a finite-size contribution, which prevents the problem of divergence at $\kappa\rightarrow0$. As already mentioned in Sec. \ref{section_fs_correction} the resulting density profiles show the incorrect behavior of a non-monotonic density profile, cf. Fig.~\ref{fig:LDA-finitesize}.

An improved correction which only partly takes edge effects into account can yet be obtained by removing the explicit spatial dependence within the density of interaction energy
\begin{equation}
\begin{split}
	u_0(\kappa,r)&=n_0\frac{N-1}{2N}Q^2\int_{\mathcal{K}(R,0)} d^3r_2\,n_0 \frac{e^{-\kappa|{\bf r}-{\bf r}_2|}}{|{\bf r}-{\bf r}_2|}\\
	&=n_0^2\frac{N-1}{N}Q^2 2\pi\int_0^R dr_2\, r_2 e^{-\kappa r_2}\\
	&=\frac{N-1}{N}Q^2\,n_0^2 \frac{2\pi}{\kappa^2}\bigl(1-e^{-\kappa R}(1+\kappa R)\bigr).
\end{split}
\end{equation}
This expression also has no divergent limit for $\kappa\rightarrow0$, and, at the same time, yields monotonically decreasing density profiles as can also be seen in Fig.~\ref{fig:LDA-finitesize}.

\end{document}